\documentclass[aps,pre,twocolumn,showpacs,superscriptaddress,noeprint]{revtex4-1}
\usepackage[utf8]{inputenc}
\usepackage{amsmath, amsfonts, amsthm, bbm}
\usepackage{amsthm}
\usepackage{amssymb}
\usepackage{pifont}
\usepackage{graphicx} 
\usepackage{bmpsize}
\usepackage{tikz}
\usepackage{float}

\begin{document}

\title{Disorder Unleashes Panic in Bitcoin Dynamics}

\author{Marco Alberto Javarone}
\thanks{These authors contributed equally to this work.\\marcojavarone@gmail.com}
 \affiliation{Centro Ricerche Enrico Fermi, Rome, Italy}
  \affiliation{University College London - Centre for Blockchain Technologies, London, UK}

\author{Gabriele Di Antonio}
\thanks{These authors contributed equally to this work.\\marcojavarone@gmail.com}
 \affiliation{Centro Ricerche Enrico Fermi, Rome, Italy} 
  \affiliation{Istituto Superiore di Sanità, Rome, Italy}
 \affiliation{Università degli Studi Roma Tre, Rome, Italy}

\author{Gianni Valerio Vinci}
\thanks{These authors contributed equally to this work.\\marcojavarone@gmail.com}
 \affiliation{Istituto Superiore di Sanità, Rome, Italy}
  \affiliation{Università Roma Tor Vergata, Rome, Italy}

\author{Raffaele Cristodaro}
 \affiliation{Blockchain and Distributed Ledger Technologies, Institute of Informatics, University of Zurich, Z{\"u}rich, Switzerland}

\author{Claudio J. Tessone}
 \affiliation{Blockchain and Distributed Ledger Technologies, Institute of Informatics, University of Zurich, Z{\"u}rich, Switzerland}
 \affiliation{UZH Blockchain Center, University of Zurich, Z{\"u}rich, Switzerland}

\author{Luciano Pietronero}
 \affiliation{Centro Ricerche Enrico Fermi, Rome, Italy}

\date{\today}
\begin{abstract}
The behaviour of Bitcoin owners is reflected in the structure and the number of bitcoin transactions encoded in the Blockchain. 
Likewise, the behaviour of Bitcoin traders is reflected in the formation of bullish and bearish trends in the crypto market. 
In light of these observations, we wonder if human behaviour underlies some relationship between the Blockchain and the crypto market. 
To address this question, we map the Blockchain to a spin-lattice problem, whose configurations form ordered and disordered patterns, representing the behaviour of Bitcoin owners. 
This novel approach allows us to obtain time series suitable to detect a causal relationship between the dynamics of the Blockchain and market trends of the Bitcoin and to find that disordered patterns in the Blockchain precede Bitcoin panic selling. 
Our results suggest that human behaviour underlying Blockchain evolution and the crypto market brings out a fascinating connection between disorder and panic in Bitcoin dynamics.
\end{abstract}
\maketitle
\textit{Introduction -} Blockchain~\cite{satoshi01,antonopoulos01} is a distributed ledger technology introduced by Satoshi Nakamoto~\cite{satoshi01}, rapidly expanding in many sectors of our society, the economy and industry.
Among the several applications, cryptocurrencies such as Bitcoin represent the most successful ones.
Bitcoin is a digital currency whose transactions get managed by a fully decentralised system that hinges on a blockchain. The latter has a data structure composed of a chain of blocks. Each block stores a set of transactions commonly verified by block creators termed \textit{miners} in this context. Within the block size limit, the miners can receive an incentive to add as many transactions as possible. 
Nevertheless, the chain of blocks keeps growing no matter the amount of executed transactions since, in principle, even blocks with no transactions can be mined and added to the chain.
Cryptographic protocols protect the Blockchain from double-spending~\cite{javarone02} and other risks. Remarkably, while fiat money requires third-party authorities, such as banks, to verify transactions, Blockchain does not need any additional authority.
Over the years, many blockchains based on new tokens, such as Bitcoin (BTC), have been implemented. These tokens are also called cryptocurrencies, or cryptos, due to the underlying cryptographic mechanisms supporting and securing transactions. 
Nowadays, a crypto ecosystem~\cite{baronchelli01,gdmarzo01,tasca01,tessone03} which includes, for instance, Ethereum (ETH), XRP (XRP), Cardano (ADA), Bitcoin Cash (BCH), Solana (SOL), Dogecoin (DOGE), Bitcoin Satoshi Vision (BSV), and many other tokens, continuously grows.
Many cryptos of such an ecosystem get exchanged in the crypto market and accessed by several trading platforms. Like in financial markets, the crypto market shows positive (i.e. bullish) and negative (i.e. bearish) trends resulting from the behaviour of traders.
In summary, the behaviour of Bitcoin users, i.e. wallet owners, traders, and so on, is relevant for the evolution of the Blockchain and the crypto market. 
But several questions remain unanswered in this complex socio-technical system: Does human behaviour underlie some relevant relationship between the Blockchain and the crypto market? The goal of this investigation is to face this question.
To this end, we map the Blockchain to a spin model, which allows assessing and measuring interactions with the crypto market.
Before moving to the details of the proposed model and related results, we remark that the Blockchain and cryptocurrencies constitute a modern and expanding research area.
Just to cite a few, previous investigations studied the Bitcoin price dynamics~\cite{kristoufek01,blau01,amjad01,aalborg01}, the crypto network of transactions~\cite{javarone01,tessone01,tessone02,tessone05,tessone06,tessone07,tessone08}, the predictive signals~\cite{yum01}, using social data~\cite{tessone04,garcia01,ortu01,marchesi02} and machine learning-based approaches~\cite{marchesi01,baronchelli02}, and the interplay between the network of Bitcoin transactions and the crypto market~\cite{kondor01}.
\newline
\newline
\textit{From Data to Model -} Datasets used in this investigation refer to a time interval from 2013 to 2022, including about $518643$ blocks and $730662636$ transactions. Blockchain data can be accessed at~\cite{blockchain_explorer} and crypto market data at~\cite{btc_trend}.
Blockchain data describe blocks and contain Bitcoin transactions and other parameters such as the Timestamp and the Blockheight. For instance, the Timestamp corresponds to the time a block gets 'mined' (i.e. generated), whereas the Blockheight identifies the position of a block along the chain.

As above-mentioned, we define a spin model by mapping blocks to vectors (see also~\cite{dai01}). In particular, we consider the following parameters: the number of transactions, the number of inputs, and the number of outputs per block.
The number of transactions per block has a self-explanatory meaning, while the other parameters, which refer to the structure of transactions~\cite{antonopoulos01}, need further details. To this end, we describe a simple transaction between Alice and Bob.
Alice owns $3$ BTC, collected from previous transactions, and wants to send $2.5$ BTC to Bob. She previously received: $1.0$ BTC, $0.35$ BTC, $0.45$ BTC, $0.9$ BTC, and $0.3$ BTC, each constituting an 'unspent transaction output' (UTXO) for a motivation later clarified.
To send $2.5$ BTC to Bob, she has to compose a transaction using a combination of UTXOs, e.g. choosing $1.0$ BTC, $0.9$ BTC, $0.45$ BTC, and $0.35$ BTC, whose summation equals $2.7$ BTC.
The chosen UTXOs constitute the inputs of the new transaction. Then, noting that the UTXO summation is greater than the amount of Bitcoin Alice wants to send to Bob, the transaction has two outputs. The first output is addressed to Bob's wallet (i.e. $2.5$ BTC), while the other is to Alice's wallet (i.e. $0.2$). These two outputs, in turn, become UTXOs that the respective receivers (i.e. Bob and Alice) can use for future transactions.
Detailed information about the microstructure of the Blockchain, i.e. the content of its blocks, can be accessed by anyone, albeit the Bitcoin owners' identity remains preserved. 
Coming back to our model, using three parameters, each block gets represented by a $3$-dimensional normalised vector \textbf{B}, and the Blockchain gets mapped to a one-dimensional lattice. The resulting structure resembles an $n$-vector model~\cite{stanley01} with $n=3$.
Now, we highlight that the content of blocks cannot change over time, as the Blockchain is an immutable ledger. Therefore, although new spin vectors add to the chain, those added in the past do not modify their configuration. Also, spin vectors forming the current chain do not affect spin vectors that will add in the future.
In summary, the Blockchain does not evolve as an Ising-like model. However, that does not prevent defining a Hamiltonian function, for instance, by fixing an instant of time to consider a limited number of spin vectors. In addition, we may assume that the spin configuration we observe at a given time represents an equilibrium configuration obtained at some temperature.
In general, the Hamiltonian of a spin model minimises at low temperatures as ordered spin patterns emerge. Similarly, it increases its value at high temperatures as disordered spin patterns show up.

As detailed below, the formation of ordered and disordered spin patterns offers valuable information to analyse the evolution of the Blockchain.
Then, the Hamiltonian of the obtained spin model (see also~\cite{welsh01}) reads
\begin{equation}\label{eq:hamiltonian_basic}
H(\textbf{B};T) = \sum_{i,j} J_{i,j} ( 1- (\textbf{B}_{i}^\intercal \textbf{B}_j))
\end{equation}
\noindent with $J_{i,j}$ interaction weight whose value is set to $0$ if $i>j$, and $\textbf{B}_{i}$ spin vector corresponding to the $i$-th block of the chain (the index $i$ represents the Blockheight and goes from $0$ to $T$).
The scalar products in equation~\ref{eq:hamiltonian_basic} get close to $1$ when consecutive vectors, i.e. blocks, are similar, otherwise get close to $0$. Note that the scalar product usually can range from $-1$ to $+1$. However, according to the range of values of the selected block parameters, the scalar product can span the interval $[0,+1]$.
Eventually, to include long-range interactions in the Hamiltonian, whose amplitude decays with the distance $J_{t-k, t} \sim e^{-\frac{k}{\tau}}$, Equation~\ref{eq:hamiltonian_basic} gets re-written as follows:
\begin{equation}\label{eq:hamiltonian_next}
H(\textbf{B}; T) = \sum_{t=0}^{T} 1 - \sum_{k=1}^{t} \frac{e^{-\frac{k}{\tau}}}{Z_t} \textbf{B}_{t-k}^{\intercal} \textbf{B}_t
\end{equation}
\noindent with $Z_t = {\sum_{k=1}^{t} e^{-\frac{k}{\tau}}}$.
In doing so, each block interacts with all previous ones. However, the exponential term weights the interactions between blocks, decaying over long distances. Such a decay gets controlled by the parameter $\tau$.
Using the single components which sum over in equation~\ref{eq:hamiltonian_next}, we obtain a collection of spin configurations that form ordered and disordered patterns.
Lastly, we emphasise that the formation of ordered and disordered patterns in the $3$-vector model can get exploited for studying the relationships between the Blockchain and trends of Bitcoin in the crypto market.
\newline
\newline
\begin{figure}
    \centering
    \includegraphics[width=\linewidth]{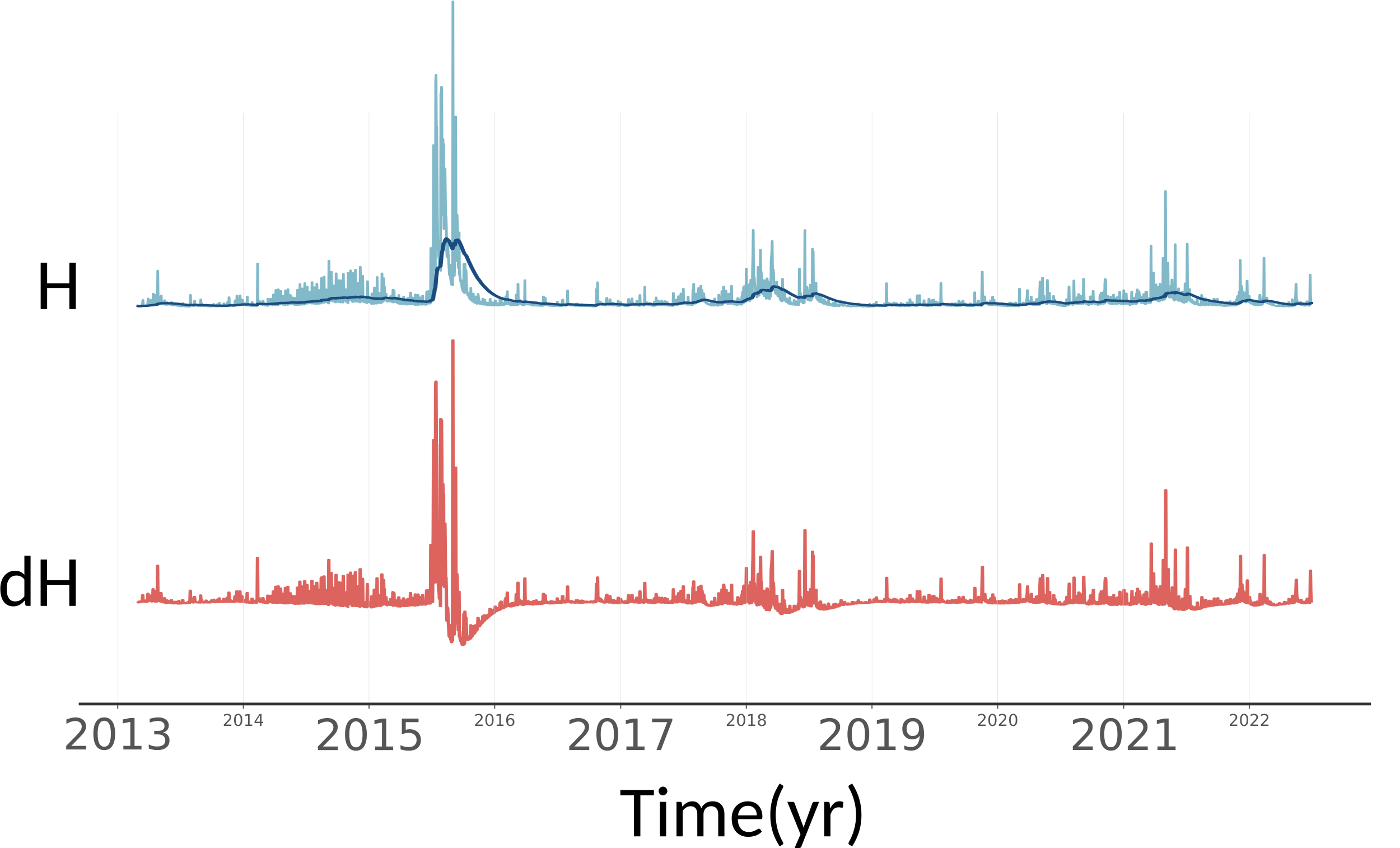}
    \caption{Blockchain evolution derived from local Hamiltonian (H) and its time derivative (dH). The small fluctuations indicate blocks are usually very similar to each other. In July 2015, a clear peak shows up.}
   \label{fig:figure_H}
\end{figure}
\textit{Results -} The Hamiltonian defined in~\ref{eq:hamiltonian_next} can be decomposed in single contributions $H(\textbf{B}; T) = \sum_{t=0}^{T} H_t(\textbf{B})$ forming a time series, to which we refer to as $H$. The latter, shown on the top of Figure~\ref{fig:figure_H}, gets computed by setting a $\tau$ small enough to include only significant long-range interactions limited to the previous $90$ days. 
\begin{figure*}
    \centering
    \includegraphics[width=\linewidth]{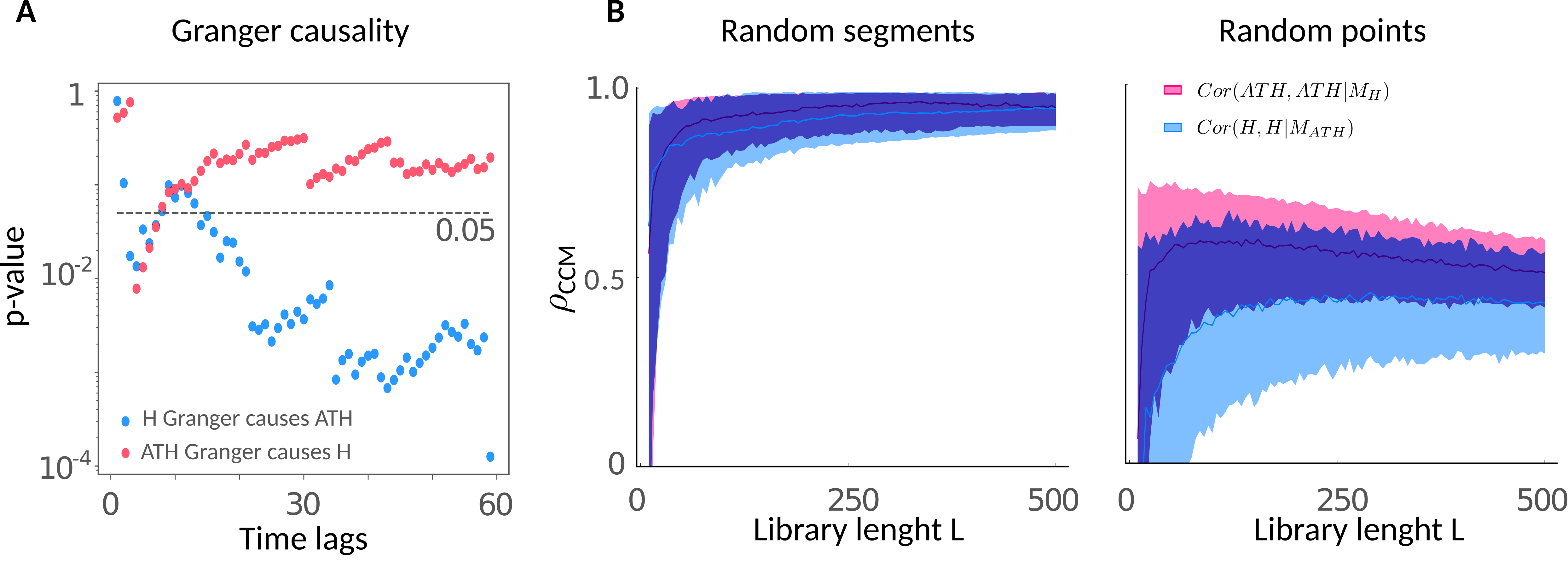}
    \caption{Causality test between $H$ and $ATH$. \textbf{A}) The p-value for Granger causality (using F-test) is shown as a function of time lags used in the fit. Increasing the knowledge of the past of $H$ in forecasting $ATH$ leads to higher significance in the test (always below p=0.05), whereas the opposite is found when trying to forecast $H$ using $ATH$. \textbf{B}) The results of the CCM test for the same variables. The convergence to an asymptotic value of the reconstruction of the variables, increasing the library length, indicates causal relation in both directions. The test was performed, for each L, using $400$ random points or contiguous segments. The straight lines are median values, and the error is computed with the $95$-th percentile.}
    \label{fig:Causality}
\end{figure*}
In addition, after applying an exponential moving average (EMA) to $H$, we compute its time derivative. The resulting time series, i.e. $dH$, is shown at the bottom of Figure~\ref{fig:figure_H}.
Interestingly, signals in Figure~\ref{fig:figure_H} have a prominent peak temporally located around July 2015. After looking for the possible sources of such a peak, we found it corresponds to the \textit{Flood} attack~\cite{flood01}, a stress test performed for testing the Bitcoin network. 
Both $H$ and $dH$ can get used for studying causal relationships with the crypto market and related phenomena. 
To this end, we focus on the $BTC/USD$ ratio (i.e. Bitcoin in American Dollars) and work on the time series composed of samples of the percentual drawdown from the All-Time-High of the $BTC/USD$ ratio.
Notably, these samples equal $1$ every time Bitcoin overcomes its previous historical maximum.
Accordingly, the time series related to the $BTC/USD$ ratio, which we refer to as $ATH$, and $H$ have the same range.
To study the causal relationship between the Blockchain and the crypto market, we use only the $H$ time series, as $H$ and $dH$ are strongly related. However, we anticipate that the $dH$ time series becomes particularly relevant in the subsequent analysis.
Accordingly, we now aim to infer a causal relationship between $H$ and $ATH$, whose task constitutes a complex and old problem~\cite{Cecconi}. For this purpose, we consider approaches relying on statistics and dynamical systems theory.
For the first case, i.e. approaches based on statistics, we perform the Granger causality test~\cite{granger1969}. Given two variables, $x$ and $y$, the Granger causality test compares the forecasting quality of future values of $y$, of a standard ARMA model (Null-hypothesis), with the same ARMA having additional information on previous values of the variable $x$. 
More in detail, $x$ is said to be Grange-cause of $y$ whether the quality of the forecasting using information over $x$ is significantly higher (p-value $<0.05$) than the quality of the forecasting obtained without $x$. 
The plot \textbf{A} in Figure~\ref{fig:Causality} shows the result of this test, whose variables are $H$ and $ATH$. In that figure, we report the p-value as a function of previous data points used in the fit of the ARMA model.
Interestingly, the quality of the fit improves as the forecasting of $ATH$ exploits more information on $H$, i.e. more historical data, suggesting a strong causal effect of $H$ over $ATH$. The reverse is not the case since, according to the p-value, we have to accept the Null-Hypothesis, i.e. $ATH$ has no Granger causality over $H$. 
For the sake of completeness, we also perform a Cross Convergent Mapping (CCM)~\cite{sugihara01,Vinci} test, which relies on dynamical systems theory and is deeply related to the Takens theorem and embedding theory~\cite{Kantz}. In this case, we look for a deterministic causality, which means that if two variables belong to the same dynamical system, one of them could be reconstructed by using the other (and vice versa) via a delayed embedding. 
Here, we consider the quality of the reconstruction of a variable $y$ through a second variable $x$, which we refer to as $y|M_x$. If such a quality increases with the number of data samples (defined as library length) the variable $x$ causally influences $y$. 
Moreover, the faster the convergence to an asymptotic value of the CCM test, the stronger the dependence between the considered variables.
A critical point of the CCM test lies in the sample selection to compose the library. So, following~\cite{Luo}, we perform the CCM test composing the library of samples by a random selection of contiguous segments and by a random selection of samples. Results are reported in the plot \textbf{B} of Figure~\ref{fig:Causality}.
Here, we observe that both sampling strategies suggest a causal relationship between the two variables, i.e. $H$ and $ATH$, in both directions. 

\begin{figure*}
    \centering
    \includegraphics[width=\linewidth]{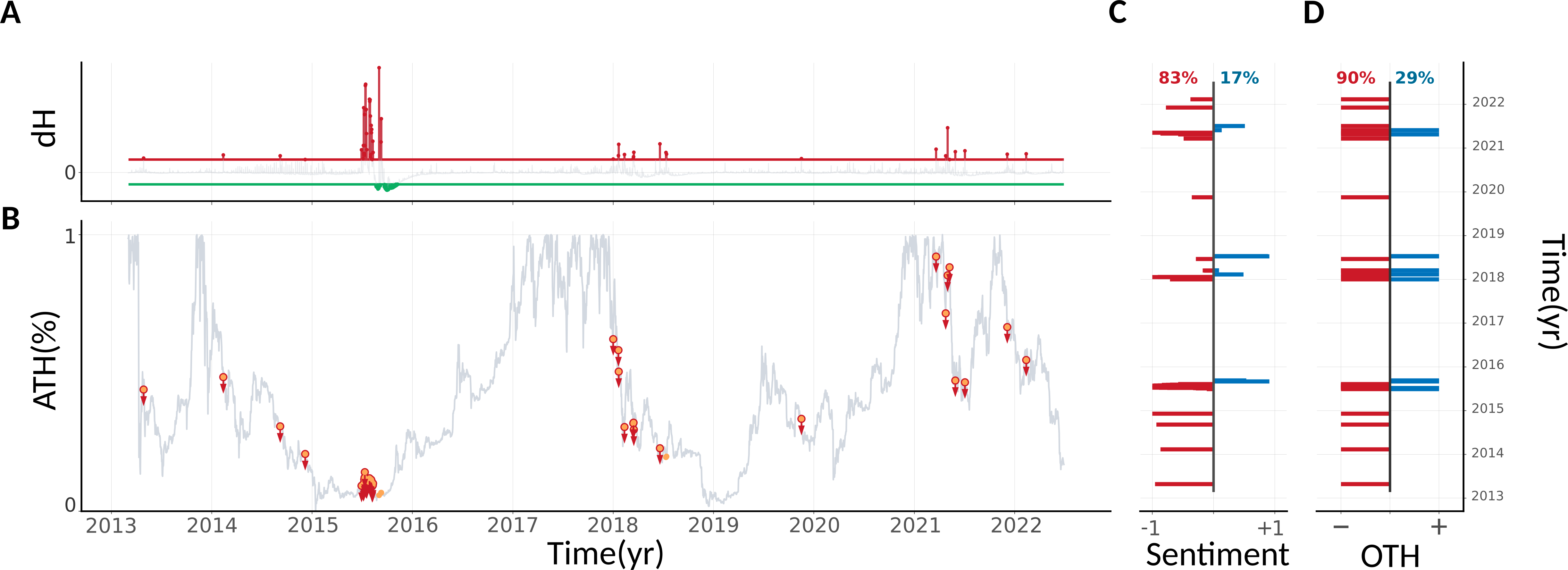}
    \caption{\textbf{A}) The trend of the $dH$ time series. Coloured lines indicate positive (red, above the $98.5$ percentile) and negative (green, below the $1.5$ percentile) fluctuations. \textbf{B}) Percentual drawdown from the $ATH$ in time. Colored dots identify events of large fluctuations in $dH$. Red arrows highlight events followed by a decrease of at least $10\%$ in the next $90$ days. Rapid increases in the $H$ time series can predict collapses in the $BTC$ price, while fast relaxation may indicate a local market recovery. \textbf{C}) Sentiment level after $dH$ positive large fluctuation. In the $83\%$ of cases, these events predate interval in which the $BTC$ price remains below the initial price most of the time (taken a $90$-day time window). \textbf{D}) Boolean of the exceeding the minimum price change threshold ($\pm10\%$, respectively $OTH=\pm$) after $dH$ positive large fluctuation. $90\%$ of these signals gets followed by a price reduction of at least $-10\%$ (in a $90$-day time window), while only the $29\%$ gets followed by a $10\%$ price growth highlighting a clear downtrend.}
   \label{fig:figure_3}
\end{figure*}
We deem that the difference between the results obtained by the CCM test and the Granger causality test, i.e. a bi-directional causal relationship and a one-directional causal relationship, respectively, might be motivated considering that the Granger causality test can only detect linear causal relationships.
In light of the above result, we study whether the $H$ time series contains information to forecast Bitcoin trends in the crypto market.
Remarkably, rapid variations of $H$ predate large fluctuations of $ATH$. Therefore, the $dH$ time series becomes particularly relevant for quantifying such phenomenon (Figure~\ref{fig:figure_3}). More in detail, we observe that rapid increases in the $H$ time series (i.e. large positive fluctuations of $dH$) can predict collapses in the $BTC$ value, while fast relaxations may indicate a local market recovery ---see plot \textbf{B} in Figure~\ref{fig:figure_3}. In addition, the $90\%$ of positive large fluctuations of $dH$ are followed by a reduction of the $BTC$ value of at least $-10\%$ (see plot \textbf{D} in Figure~\ref{fig:figure_3}).
\newline
\newline
\textit{Conclusion -} In summary, this work unveils relevant relationships between the dynamics of the Blockchain and the crypto market, focusing on the Bitcoin price. The investigation, motivated by observing that human behaviour affects both the dynamics of the Blockchain and those of the crypto market, exploits tools from statistical physics.
More specifically, we generated time series describing the evolution of the Blockchain via a spin-lattice model. Such time series allowed us to obtain the following results. Firstly, we detected a causal relationship between the Blockchain and the crypto market, and then we found Blockchain contains information to forecast some trends in Bitcoin price.
Remarkably, disordered patterns in the Blockchain, identified via the spin model, predate the phenomenon of Bitcoin panic selling, suggesting a fascinating connection between disorder and panic.
Before concluding, let us report a few observations about some previous investigations.
In~\cite{aalborg01}, authors highlight the potential role of Bitcoin transactions in driving the Bitcoin trading volume and price. That is confirmed by our results, as we show that the number of Bitcoin transactions plays a role in forecasting the Bitcoin market trends.
In addition, some ideas and outcomes of our investigation remind works~\cite{preis01,preis02}, which aimed at forecasting financial market trends by looking at Wikipedia and Google Trends analytics, respectively. Likewise, here we aim to foresee relevant phenomena in the crypto market, e.g. panic selling, by exploiting analytics data related to an external system, i.e. the Blockchain.
Finally, we deem the proposed model sheds light on relevant aspects of Bitcoin dynamics. Therefore, future works based on this investigation could address the behaviour of other cryptocurrencies and assess whether related results can support the design of trading strategies for the crypto market.
\section*{Acknowledgement}
MAJ wishes to thank Marco Corradino for stimulating discussions and Mario Bortoli for helpful suggestions.

\end{document}